\documentclass[twocolumn,showpacs,amsfonts,amssymb,floats,aps]{revtex4-1}

\usepackage{graphicx}

\def\Tr#1{\textrm{Tr}\left[#1\right]}

\def\ket#1{|#1\rangle }

\def\lph{\lambda_{ph}}
\def\w{\omega}
\def\e{\varepsilon}

\def\non{\nonumber \\ }

\begin{document}

\title{Quantum transport through a molecular level:
a scattering states numerical renormalisation group study}

\author{Andre Jovchev}
\author{Frithjof B. Anders}
\address{Lehrstuhl f\"ur Theoretische Physik II, Technische Universit\"at Dortmund
Otto-Hahn-Str. 4, 44221 Dortmund, Germany
}

%
%
\begin{abstract}
We use the scattering states numerical renormalization group (SNRG)
approach to calculate the current $I(V)$ through a single molecular
level coupled to a local molecular phonon. The suppression of $I$ for
asymmetric junctions with increasing electron-phonon coupling, the
hallmark of the Franck-Condon blockade, is discussed.  We compare the
SNRG currents with recently published data obtained by an iterative
summation of path integrals approach (ISPI).  Our results excellently
agree with the ISPI currents for small and intermediate voltages. In
the linear response regime $I(V)$ approaches the current calculated
from the equilibrium spectral function.  We also present the
temperature and voltage evolution of the non-equilibrium spectral
functions for a particle-hole asymmetric junction with symmetric
coupling to the lead.

\end{abstract}
\maketitle

\section{Introduction}

In the quest for size-reduced and possible low-power consuming
electronic devices, the proposal \cite{AviramRatner1974}  of using
molecular junctions for electronics has sparked a  large interest in
understanding the influence of molecular vibrational modes onto the
electron charge transfer through a molecule.  
The non-linear current through a  molecule device can be controlled
by a capacitively coupled external gate \cite{Chen1999,Donhauser2001}. 
Interestingly, hysteretic behavior of the $I(V)$
curves \cite{LiHysteresis2003} has been  reported in several experiments when sweeping the
voltage with a finite rate. However, the observed hysteresis are non-universal and
depend on the sweeping rate. For infinitesimally slow sweeping the effect vanishes.
In some cases a sudden drop of
the current has been observed  with increasing bias
voltage \cite{Chen1999} which translates into a negative differential
conductance. This all has been accounted to configural changes of the molecule
emphasizing the importance of vibrational couplings in such devices.

Many experimental facts have been gathered in the last two decades but
there is still a lack of an accurate theoretical description of all
the reported phenomena. An excellent review
\cite{GalperinRatnerNitzan2007} by Galperin et al.\ comprehensively
summarizes the different theoretical approaches and experimental
findings.  Single molecular transistors (SMT) promise to offer some
advantages over their semi-conductor based counterparts
\cite{KastnerSET1992}. Both types of single-electron transistors can
be controlled by a capacitively coupled external gate
\cite{KastnerSET1992,Chen1999,Donhauser2001}.  The molecular energy
scales, however, are larger in SMTs and reproducibly defined by the
chemistry of the molecule. In addition, the coupling to vibrational
modes enlarges the parameter space and different physics such a
phonon-assisted tunneling, Frank-Condon blockade or the appearance of
inelastic steps in the $I(V)$ curve can be observed.

The theoretical description of such molecular junctions only include
those molecular levels and vibrationals modes relevant for the quantum
transport. The simplest model proposed
\cite{GalperinRatnerNitzanHysteresis2004,GalperinRatnerNitzan2004,GalperinRatnerNitzan2007}
comprise a single level coupled to a local Holstein phonon.  Typically
rate equations \cite{KochOppen2005} or lowest order Keldysh-Green
function approaches
\cite{GalperinRatnerNitzan2007,KochFehske2011,KFLFQMT11} have been
applied to this problem \cite{GalperinRatnerNitzanHysteresis2004}.
Recently, the iterative path-integral approach (ISPI)
\cite{weissthorwart2008} has also been successfully applied
\cite{HutzenEgger2012} to calculate quantum transport for moderate and
high temperatures compared to the charge-transfer rate $\Gamma_0$.

The equilibrium physics of two extreme limits have been well
understood in a model containing only a single vibrational mode
\cite{GalperinRatnerNitzan2007,KochOppen2005,EidelsteinSchiller2012}.
In the adiabatic limit, the phonon frequency is the smallest energy
scale of the problem and a small electron-phonon coupling yields a
reduction of the phonon frequency by particle-hole excitations.  This
limit has been pioneered by Caroli et al.\cite{Caroli71,*Caroli72} in
the context of tunnel junctions and applied to molecular junctions
\cite{GalperinRatnerNitzan2004} .

In the opposite limit, for very small tunneling rates $t_\alpha$ one
starts from the exact solution of the local problem by applying a
Lang-Firsov transformation \cite{LangFirsov1962}.  A displaced phonon
with an unrenormalized phonon frequency $\w_0$ and a polaron with a
shifted single-particle energy is formed locally.  In this
anti-adiabatic limit, the strong electron-phonon coupling yields a
polaronic shift of the single-particle level and a exponential
suppression of tunneling rate related to the Franck-Condon blockade
\cite{KochOppen2005,GalperinRatnerNitzan2007,NatureFranckCondonCarbonNanotubes2009,HutzenEgger2012}.

Wilson's numerical renormalization group (NRG) approach
\cite{Wilson75,BullaCostiPruschke2008} has been adapted to the
Holstein model in equilibrium \cite{HewsonMeyer02}.  A comprehensive
study \cite{EidelsteinSchiller2012} has demonstrated the power of this
non-perturbative approach to reveal the interplay between the
different energy scales of the problem in the crossover regime.  In
this article we review the extension \cite{JovchevAnders2013} of the
approach to steady-state currents by applying the scattering-states
NRG (SNRG) \cite{AndersSSnrg2008,SchmittAnders2010,SchmittAnders2011}
to the spinless Anderson-Holstein model.

\section{Theory of Quantum transport through molecular junction}

\subsection{Model}

In molecular electronics experiments \cite{Chen1999,Donhauser2001}, a
complex organic molecule is contacted by two conducting leads. We have
modeled these leads as two symmetric featureless free electron gases
since the mean-free path in the leads is large compared to the spatial
dimensions of the device. In general, the molecule can contain several
molecular orbitals which are actively participating in the quantum
transport. Furthermore, the internal vibrational modes of the molecule
are influenced by charging and discharging of the molecule.

\begin{figure}[t]
\begin{center}
\includegraphics[width=65mm]{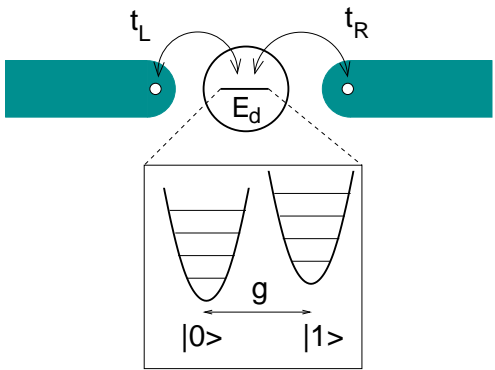}

\caption{Minimal model of a molecule consisting of a single active
molecular level at energy $E_d$ coupled to two leads with tunneling
matrix elements $t_L$ and $t_R$.  Depending on the local charge
configuration $\ket{0}$ or $\ket{1}$, the ground state of a
vibrational degree of freedom is shifted. The relative displacement
between the two configuration is given by the dimensionless electron
phonon coupling $g=\lph/\w_0$.  The phononic excitations for a fixed
charge are multiples of the oscillator energy $\w_0$.
}
\label{fig:1-model}
\end{center}
\end{figure}

Here we consider the most minimalistic model for quantum transport
through a molecule
\cite{KochOppen2005,GalperinRatnerNitzan2007,HutzenEgger2012} and
follow the notation of Ref.\ \cite{JovchevAnders2013}.  This widely
used Hamiltonian
\cite{GalperinRatnerNitzan2007,HutzenEgger2012,EidelsteinSchiller2012}
is defined as
\begin{eqnarray}
\label{equ:H}
H &=&  E_d d^\dagger d +\w_0 b^\dagger b + \lambda_{ph}  (b^\dagger + b)\left( \hat n_d - \frac{1}{2}\right) 
\nonumber
\\
&&+ \sum_{\alpha=L,R} \sum_{k} \e_{k\alpha}c^\dagger_{k\alpha}c_{k\alpha}
\non
&& 
 +
\sum_{\alpha=L,R} \frac{t_\alpha}{\sqrt{N}} \sum_{k} (d^\dagger c_{k\alpha} + c^\dagger_{k\alpha} d)
\end{eqnarray}
where $d(d^\dagger)$ annihilates(creates) an electron on the device with
energy $E_d$, and $c^\dagger_{k\alpha}$ creates an electron in the
lead $\alpha$  with energy $\e_{k\alpha}$.  The local charge-transfer
rate   to each lead $\alpha$  is given by $\Gamma_\alpha=\pi
t_\alpha^2 \rho_\alpha(0)$, where $\rho_\alpha(\w)$ is  the  density
of states of lead $\alpha$.  In order to focus only on the influence
of the electron-phonon interaction onto the quantum transport,  the
spin degree of freedom is neglected 
in order to avoid obstruction of the competition between
spin-flip scattering through the device and polaron formation on the device.

The model comprise a single active molecular level - all others are
energetically well separated -- whose charge density is coupled to a
local Holstein phonon stemming from the dominating vibrational mode of
the molecule.  In real materials, band features are important but only
influence the single-particle properties which can be accounted for in
a frequency dependent charge transfer rate $\Gamma_\alpha(\w)$ which
we treat as a constant for simplicity in our simulations.

The spinless Anderson-Holstein model is schematically depicted in
Fig.\ \ref{fig:1-model}. Depending on the local charge configuration,
the local harmonic oscillator is displaced and the dimensionless
distance between the two ground states is given by $g=\lph/\w_0$.  For
modeling realistic situations, the restriction to a single phonon and
a single electronic level must be lifted.  In spite of a lot of
theoretical progress \cite{GalperinRatnerNitzan2007} this model has
only been accurately solved in equilibrium
\cite{HewsonMeyer02,VinklerSchillerAndrei2012,EidelsteinSchiller2012},
while its non-equilibrium dynamics has only be perturbatively
investigated in lowest order of the coupling constants
\cite{GalperinRatnerNitzan2007}.

The local Hamiltonian is given by the first line in (\ref{equ:H}) and
can be solved exactly using the Lang-Firsov transformation
\cite{LangFirsov1962,Mahan81}.  This local solution consists of a
local polaron decoupled from a shifted harmonic oscillator.  The
corresponding polaronic energy gain is given by $E_p=\lph^2/\w_0 =
g^2\w_0$.

\begin{figure}[tb]
\begin{center}
\includegraphics[width=90mm]{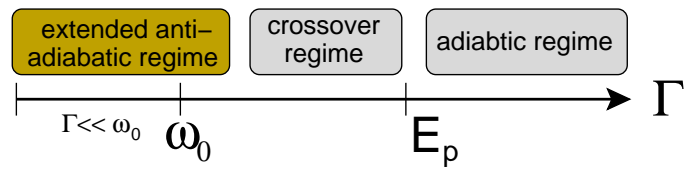}
\caption{
The different regimes as function of the charge transfer rate $\Gamma_{0}$. 
The crossover from extended
anti-adiabatic regime is reached when $\Gamma_{\rm eff} \approx e^{-g^2} \Gamma_0\approx \w_0$
and extents to $\Gamma\approx E_p = g^2\w_0$.
}
\label{fig:2-diff-regimes}
\end{center}
\end{figure}

Coupling this local degrees of freedom to the two leads defines two
competing regimes depicted in Fig.\ \ref{fig:2-diff-regimes}.  For
$E_p,\w_0 \ll \Gamma_0=\Gamma_L+\Gamma_R$, the phonon dynamics is slow
and can be treated perturbatively in this adiabatic
regime. $E_p,\w_0\gg \Gamma_0$ defines the opposite limit: in this
anti-adiabatic regime charge fluctuations are suppressed, the electron
moves slow and the phonon defines the large energy scale. The
anti-adiabatic regime is relevant for molecular junctions since the
tunneling coupling of a molecule to the leads is usually small
compared to the intrinsic energy scales of the molecule. After the
Lang-Firsov transformation, the tunneling term acquires an additional
factor $\exp[g(b^\dagger -b)]$ whose physical meaning is stripping the
original electron content from the locally formed polaron. If $\w_0\gg
\Gamma$, the local phonon remains in its ground states which yields an
exponential suppression of the tunneling coupling and $\Gamma\to
\Gamma_{\rm eff}\approx \Gamma_0 e^{-g^2}$.  In a particle-hole
asymmetric junction, this leads to a Franck-Condon suppression of the
current for small bias voltage: The system reacts with a dynamical
suppression of the tunneling rate to avoid the reorganization of the
nuclear positions of the molecule.

\subsection{Scattering-states numerical renormalization group}
\label{sec:snrg}

The scattering-states numerical renormalization group (SNRG) approach
is based on the steady-state density operator $\rho(\mu_L, \mu_R)$ for
a current carrying ensemble \cite{Hershfield1993} coupled to two baths
at different chemical potentials $\mu_\alpha$. Hershfield has shown
\cite{Hershfield1993} that this operator has a Boltzmannian form
\begin{eqnarray}
\hat \rho(\mu_L, \mu_R) &=& \frac{1}{Z}e^{-\beta(H-Y)}
\end{eqnarray}
where $Z$ is the partition function and $\mu \hat N$ is replaced by
the $Y$-operator.  This $Y$-operator is in general unknown for a
arbitrary fully interacting Hamiltonian.

For a non-interacting problem, however, the corresponding
$Y_0$-operator is given in terms of the Lippmann-Schwinger scattering
states with energy $\e$ of left-moving or
right moving single-particle scatting state created by $\gamma^\dagger_{\alpha}(\e)$,
\begin{eqnarray}
Y_0 &=& \sum_\alpha \mu_\alpha \int d\e \, \gamma^\dagger_{\e \alpha}\gamma_{\e \alpha}
\end{eqnarray}
where $\{\gamma^\dagger_{\e \alpha},\gamma^\dagger_{\e' \alpha'} \}
=\delta_{\alpha \alpha'}\delta(\e-\e')$.  Note that the applied
source-drain voltage $eV=\mu_L-\mu_R$.

In the SNRG \cite{AndersSSnrg2008,SchmittAnders2010} we have
circumvented the unknown $Y$ by the following procedure.

First, we realise that we can discretize the energy dependent
scattering states on a logarithmic energy mesh identically as in the
standard NRG\cite{BullaCostiPruschke2008,AndersSSnrg2008} such that we
obtain a two-band model comprising of a left-mover and right-mover
band. (Below, we will comment on the analytical form of the scattering
states.) Then we perform a standard NRG using $K_0=H(\lph=0) -\hat
Y_0$.

Knowing the analytical form of the non-equilibrium density operator
$\hat \rho_0(V)=\exp(-\beta K_0)/Z_0$, we can discretize scattering
states on a logarithmic energy mesh identically to the standard NRG
\cite{BullaCostiPruschke2008,AndersSSnrg2008} and perform an NRG using
$K_0=H(\lph=0) -\hat Y_0$. The density operator $ \hat \rho_0(V)$
contains all information about the current carrying steady-state for
the Hamiltonian $H_0=H(\lph=0)$ \cite{Hershfield1993}.

Starting at time $t=0$, we let the non-interacting system propagate
with respect to the full Hamiltonian $H_f=H(\lph>0)$: The density
operator $\hat \rho(t)$ progresses as $\hat \rho(t) = \exp(-iH_f t)
\rho_0 \exp(iH_f t)$.  Since we quench the system only locally, we can
assume $\hat \rho(t)$ reaches a steady-state at $t\to\infty$
independent of initial condition for an infinitely large system: all
bath correlation functions decay for infinitely long times.  The
finite size oscillations always present in the NRG calculation
\cite{AndersSchiller2005,AndersSchiller2006,AndersSSnrg2008,EidelsteinGuettgeSchillerAnders2012,GuettgeAndersSchiller2013}
are projected out by defining the time-averaged density operator
\cite{Suzuki11971,AndersSSnrg2008}
\begin{eqnarray}
\hat \rho_\infty &=& \lim_{T\to\infty} \frac{1}{T}\int_0^{T} dt \hat \rho(t)
\, .
\label{eq:rho-infity}
\end{eqnarray}
Consequently, only density matrix elements diagonal in energy contribute to the steady-state
in accordance  with the condition  $[H_f,\hat \rho_{\infty}] =0$.  Even though the $Y$-operator
remains unknown, we explicitly construct a numerical representation of the non-equilibirum
density matrix using the 
time-dependent NRG \cite{AndersSchiller2005,*AndersSchiller2006,EidelsteinGuettgeSchillerAnders2012}.
In a last step, we calculate local steady-state retarded Green function
\begin{eqnarray} 
\label{eq:neqGf}  G^r_{d,d^\dagger}(t) &= & 
- i \Tr{ \hat \rho_{\infty} \{  d(t), d^\dagger \} } \Theta(t), \label{eqn:steady-state-gf}   
\end{eqnarray}
where $d(t)=e^{i H_f t} d e^{-i H_f t}$ and  $\hat \rho_{\infty}$ has been defined in
Eq.\ (\ref{eq:rho-infity}). 
The approach is based on an extension for equilibrium Green
functions \cite{PetersPruschkeAnders2006} and its
technical details  are found in Ref.\ \onlinecite{AndersNeqGf2008}.

It has been 
show \cite{MeirWingreen1992,Hershfield1993,Oguri2007} that
for the model investigated here, the current
is given by the  by a generalized Landauer formula 
\begin{eqnarray}
  \label{eq:ss-current}
  I(V) &=& \frac{G_0}{e}  \int_{-\infty}^\infty \, d\w \, 
\left[f_R(\w)-f_L(\w)\right] 
\Gamma_0 \pi \rho_d(\w,V)
\end{eqnarray}
where $f_\alpha(\w)=f(\w-\mu_\alpha)$
and the steady-state spectral function
$\pi \rho_d(\w,V) = \Im m [G^r_{d,d^\dagger}(\w-i0^+,V)]/\pi$ 
is obtain from the Fourier transformed retarded Green function Eq.~(\ref{eq:neqGf}).
The prefactor
\begin{eqnarray}
G_0 &=&   \frac{e^2}{h} \frac{4\Gamma_L
\Gamma_R}{\Gamma_0^2}
\label{eqn:G0-prefactor}
\end{eqnarray}
contains the leading asymmetry factors  of the junction and 
reaches the universal conductance quantum $e^2/h$ for a symmetric 
junction, i.\ e.\ $\Gamma_L=\Gamma_R$. 
$G_0$ can be expressed as
$G_0 = ( e^2/h)(4 R/(1+R)^2)$ using the definition of the coupling
asymmetry ratio $R=\Gamma_L/\Gamma_R$.

\subsection{Single-particle scattering states}

In the absence of the electron-phonon coupling, the Hamiltonian (\ref{equ:H})
can be solved exactly in
terms of single-particle Lippmann-Schwinger scattering states \cite{Hershfield1993,SchillerHershfield95,Oguri2007,AndersSSnrg2008},
\begin{eqnarray}
H_0&=& \sum_{\alpha} \int d\e\,  \e\,  \gamma_{\e \alpha}^\dagger  \gamma_{\e \alpha}
\end{eqnarray}
where 
\begin{eqnarray}
  \label{eq:scattering-states-operators}
  \gamma^\dagger_{ \alpha} &=& c^\dagger_{\e \alpha} +
  t_\alpha \sqrt{\rho_\alpha(\e)} G_{0\sigma}^r(\e +i\delta)
 \nonumber \\ && \times 
 \left[
 d^\dagger +\sum_{\alpha'} 
 \int d\e' 
 \frac{V_{\alpha'} \sqrt{\rho_{\alpha'(\e')}}}{\e+i\delta -\e'}
 c^\dagger_{\e'\alpha'} 
 \right]
\end{eqnarray} 
and  the local resonant level Green
function 
\begin{eqnarray}
G_{0}^r(z) &=& \left[z-  E_d  - \Delta(z)
\right]^{-1}
\\
\Delta(\w-i\delta) &=& \sum_\alpha t^2_\alpha \int d\e \frac{\rho_\alpha(\e)}{\w-\e}
\end{eqnarray}
enters as one of the expansion coefficients. $\rho_\alpha(\e)$ denotes
the density of states of the individual leads and will be takes as
equal and featureless in the following.  The small imaginary part
$i\delta$ is required for regularization in the transition from the
discrete $k$ summation in Eq.\ (\ref{equ:H}) to the continuum limit an
caused the time-reversal symmetry breaking.

Note that the $d$-orbital has been included into the scattering states. By inverting the unitary
transformation, we can expand the local $d$-orbital in 
left-moving and right-moving  scattering states.
\begin{eqnarray}
d^\dagger &=& r_R d^\dagger_{R} + r_L d^\dagger_{L}
\\
\label{eqn:d-alpha-cont}
 d^\dagger_{\alpha} &=& \bar t \int d\e \sqrt{\rho (\e)} [G_{0}^r(\e +i\delta)]^* 
 \gamma ^\dagger_{\e \alpha}
\end{eqnarray}
where we defined $\bar t = \sqrt{t_L^2 +t_R^2}$  and have used $r_{\alpha} = t_{\alpha}/\bar t $.
The  expansion coefficients in Eq.\ (\ref{eqn:d-alpha-cont}) contain the  retarded Green 
function $ G_{0}^r(\e +i\delta)$ which we separate in modulus and phase
\begin{eqnarray}
 [G_{0}^r(\e +i\delta)] &=& | G_{0}^r(\e +i\delta)| e^{-i\Phi(\e)}
 .
\end{eqnarray}
This phase is absorbed into the  new scattering states  $ \gamma^\dagger_{\e \alpha} \to  \tilde \gamma^\dagger_{\e  \alpha}  =  \gamma^\dagger_{\e  \alpha}  e^{i\Phi(\e)}$ 
by a local gauge transformation.  In the wide band limit, i.e.\ $D\gg \Gamma_0$,
the effective DOS $\tilde \rho(\e)= [\bar t \sqrt{\rho(\e) } | G_{0}^r(\e -i\delta)| ]^2$
is normalized, 
\begin{eqnarray}
\int d\e \tilde \rho(\e)&=&
\int d\e \left[ \bar V  \sqrt{\rho (\e)} | G_{0}^r(\e -i\delta)| \right]^2 = 1 
\, ,
\end{eqnarray}
and $d^\dagger_\alpha$ is used as a starting vector for the
Householder transformation \cite{Wilson75,BullaCostiPruschke2008} for
constructing the discretized Wilson chain. Although the physical
contained of the Wilson chain sites are different to the standard NRG
\cite{Wilson75,BullaCostiPruschke2008} the analytical from is
preserved \cite{AndersSSnrg2008}. Since the local gauge transformation
has to applied to the local current operator, the current flow is
related to $\sin(\Phi(\e))$ of the energy dependent scattering phase.

\section{Results}

\begin{figure}[t]
\begin{center}
\includegraphics[width=90mm]{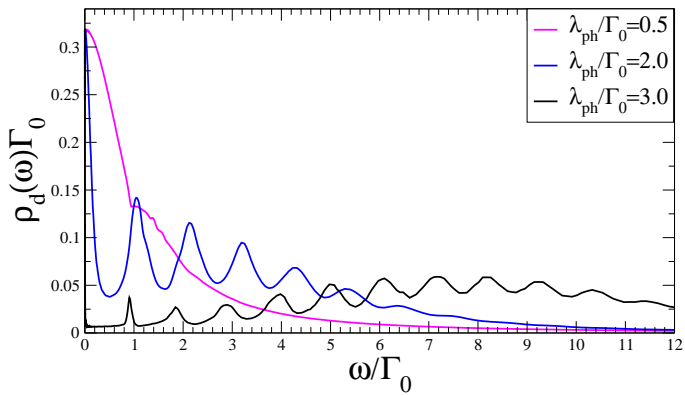}

\caption{
Equilibrium spectral function of the particle-hole symmetric junction, i.\ e.\ $E_d=0$,
and $\w_0/\Gamma_0=1$ for three different electron-phonon coupling strength $\lph/\Gamma_0=0.5,2,3$.
}
\label{fig:3}
\end{center}
\end{figure}

\subsection{Equilibrium spectral function}

To set the stage for the non-equilibrum steady state currents,
we show the evolution of the equilibrium spectra function $\rho_d(\w)$ for three different 
ratios $\lph/\Gamma$ in Fig.\ \ref{fig:3}. While $\lph/\Gamma=0.5$ lies in the adiabatic regime, the two others 
are represents the anti-adiabatic regime. For $\lph/\Gamma=0.5$, we find a kink in the spectral function at $\w_0$
where strong electron-phonon scattering sets in. For $\lph/\Gamma=2$ we  observe already very 
pronounced phonon-replicas with a reduced width. Increasing $\lph/\Gamma$ further yields to a substantial
shift of spectral weight from the resonance at $\w=0$ to larger frequencies: a careful analysis shows \cite{EidelsteinSchiller2012,JovchevAnders2013} that the peak of the envelope function
is related to the effective Coulomb repulsion between the $d$-electron and the conduction band electrons
which can be derived analytically for $t_\alpha\to 0$ using a Schrieffer-Wulff transformation \cite{EidelsteinSchiller2012}.

\subsection{Steady-state currents}

\begin{figure}[t]
\begin{center}
\includegraphics[width=90mm]{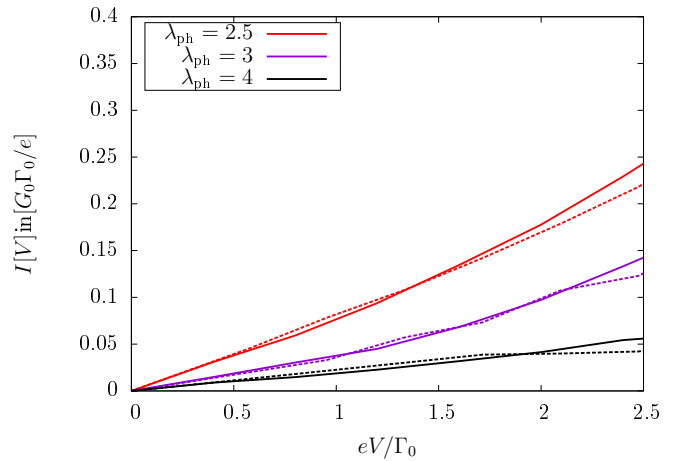}

\caption{
The evolution of $I(V)$ curves from medium to strong electron-phonon coupling for 
$T/\Gamma_0=0.2$  for a fixed phonon frequency $\w_0/\Gamma_0=2$, a symmetric
junction $\Gamma_L=\Gamma_R$ and
level position in $H_m$, i.\ e.\ $\e=E_d+\lph^2/\w_0=0$.
A comparison between the SNRG data (straight lines) and
the ISPI data (dotted lines), taken with permission from Ref.\ \cite{HutzenEgger2012},
is shown.
}
\label{fig4-ISPI-SNRG}
\end{center}
\end{figure}


Recently, a numerical approach based on the iterative summation of
path integrals (ISPI) \cite{weissthorwart2008} has been applied
\cite{HutzenEgger2012} to the model defined in Eq.\ (\ref{equ:H}).
Since it requires a fast decay of the memory kernel for the discetized
iterative summation of the path integral, it is restricted to moderate
and high temperatures for large electron-phonon couplings.  In this
section, we will provide a comparison of the ISPI with the SNRG using
the published ISPI data of Ref.\ \cite{HutzenEgger2012}.

It is straight forward to show that local Hamiltonian 
\begin{eqnarray}
\label{equ:Hm}
H_m &=&  \e d^\dagger d +\w_0 b^\dagger b + \lambda_{ph}  (b^\dagger + b) \hat n_d  
\end{eqnarray}
commonly used in the literature \cite{GalperinRatnerNitzanHysteresis2004,HutzenEgger2012}
yield the same dynamics as the first three terms in Eq.\ (\ref{equ:H}) after identifying 
$\e=E_d +g^2\w_0$ and performing a linear shift of the bosonic operators \cite{EidelsteinSchiller2012,JovchevAnders2013}.

Figure \ref{fig4-ISPI-SNRG} shows the evolution of the current for a
symmetric junction $\Gamma_L/\Gamma_R=1$ from medium to strong
coupling at $T/\Gamma_0=0.2$, $\e=0$ and $\w_0/\Gamma_0=2$.  The
overall agreement between the SNRG data (solid lines) and the ISPI
approach (dotted lines) is remarkable up to $eV\approx 2\Gamma_0$
after which the small deviations become more pronounced: The SNRG
current slightly exceeds the ISPI data for large voltages. Since both
approaches, the ISPI and the SNRG, relay on discretisation of a
continuum, we believe that the origin of these deviations are related
to the different discretisation errors inherent in both methods.

\begin{figure}[bt]
\begin{center}
\includegraphics[width=90mm]{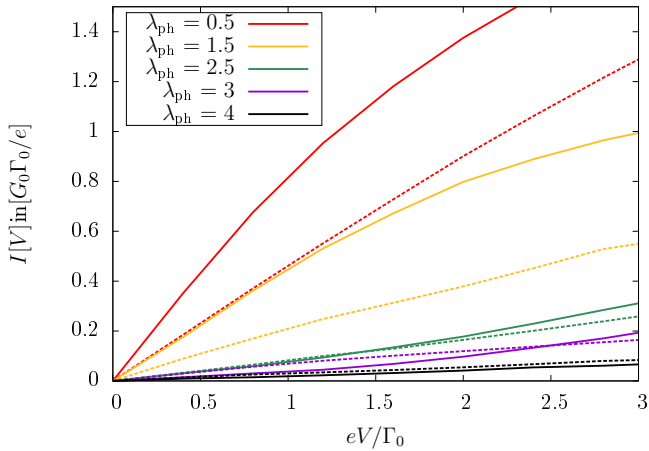}

\caption{
The evolution of $I(V)$ curves from weak to strong electron-phonon coupling for 
$T/\Gamma_0=1$ (dashed lines) and $T/\Gamma_0=0.2$ (straight lines) calculated with 
the SNRG. We have set $\w_0/\Gamma_0=2$, a symmetric
junction $\Gamma_L=\Gamma_R$ and
$\e=E_d+\lph^2/\w_0=0$.
}
\label{fig5-SNRG-temperatures}
\end{center}
\end{figure}

The temperature dependency of the SNRG $I(V)$ curves are shown in
Fig.\ \ref{fig5-SNRG-temperatures} for two different temperatures.  We
combine the data of Fig.\ \ref{fig4-ISPI-SNRG} for $T/\Gamma_0 = 0.2$
(straight lines) with the $I(V)$ for the same parameters but
calculated at $T/\Gamma_0 = 1$ (dashed lines).  In the limit $T
\rightarrow \infty$ the currents must vanish: In this high temperature
limit all left and right moving scattering states are equally occupied
leading to zero net current as predicted by Eq.\
(\ref{eq:ss-current}).
For the electron-phonon couplings $
\lambda_{\text{ph}}/\Gamma_0=0.5,1.5,2.5$, we clearly observe a
decrease of the current with increasing temperature.

Above $\lambda_{\text{ph}}/\Gamma_0=3$, we observe a qualitative
change of the behavior: the low temperature current is smaller than
its high temperature counterpart: an indication of the Franck-Condon
blockade in the quantum transport.
There are two contributions changing the current according to Eq.\
(\ref{eq:ss-current}) for a fixed voltage when raising the
temperature.  Firstly, the Fermi window becomes flatter and broader,
and the high energy parts of the spectral function contribute
stronger.  In addition, the spectral function shows a significant
temperature and voltage dependency with increasing electron-phonon
coupling.

\begin{figure}[t]
\begin{center}
\includegraphics[width=90mm]{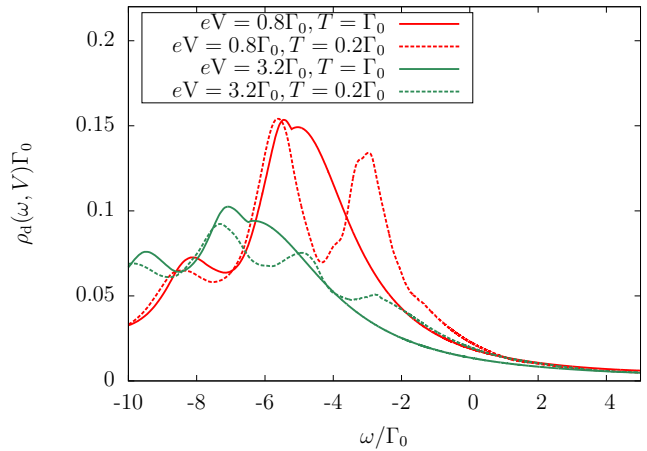}

\caption{
SNRG nonequilibrium spectral functions for two different voltages and two temperatures.
We have set $\w_0/\Gamma_0=2, \lph/\Gamma_0=3$ and
$\e=E_d+\lph^2/\w_0=0$.
}
\label{fig6:spec-nonequi}
\end{center}
\end{figure}

To illustrate this points, a comparison of nonequilibrium spectral
functions with $\lambda_{\text{ph}} = 3$ is depicted in Fig.\
\ref{fig6:spec-nonequi} for two different temperatures and two
different bias voltages.  For $eV = 0.8 \Gamma_0$ an increase of
temperature leads to a suppression of the phonon side peak at
$\w/\Gamma_0 = -3$.

At the same time the peak at $\w/\Gamma_0 = -5$ is broadened and
contributes more weight to the integral due to the broadened Fermi
window, leading to an increase of the current with increasing
temperature.
At a voltage of $eV = 3.2 \Gamma_0 $ the decrease of the spectral
weight at $\w/\Gamma_0 = -3$ is not compensated within the Fermi
window contributing to the current integral leading to a decrease of
the current with increasing temperatures.
Therefore, we observe a crossover between an increase of current at
small voltages to an decrease of current at large voltages with
increasing temperatures.

In contrary, the difference between the I-V curves of $T/\Gamma_0 = 1$
and $T/\Gamma_0= 0.2$ are large at low phonon-couplings $
\lambda_{\text{ph}}/\Gamma_0=0.5,1.5$.  In this perturbative regime,
the spectral function is only very weakly temperature dependent, and,
therefore, the change of currents is only related to the temperature
dependence of the Fermi functions in Eq.\ (\ref{eq:ss-current}).

\subsection{Linear response regime}

\begin{figure}[t]
\begin{center}
\includegraphics[width=90mm]{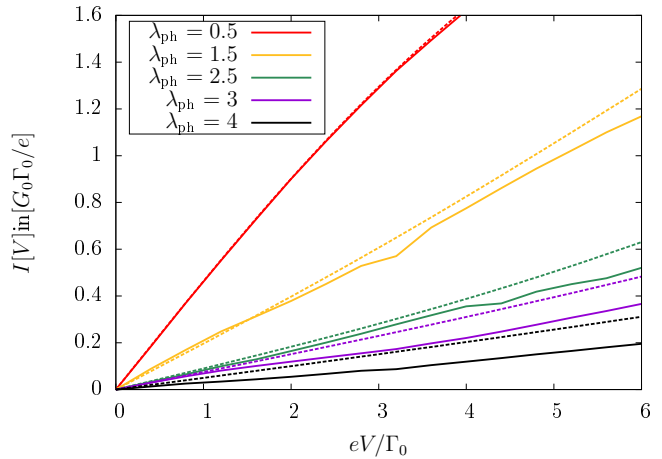}

\caption{
The evolution of $I(V)$ curves from weak to strong electron-phonon coupling for the same parameters as in fig. \ref{fig4-ISPI-SNRG} but
$T/\Gamma_0=1$. The straight lines are SNRG nonequilibrium results and the dashed lines are I(V) curves where we set $\rho_d(\w, V) = \rho_d(\w)$ in Eq. \ref{eq:ss-current}.
}
\label{fig7:equi-nonequi}
\end{center}
\end{figure}

In order to explicit reveal the influence of the voltage dependency of
the spectral function on the current $I(V)$, we compare the SNRG
curves with the current calculated with the equilibrium spectral
function $\rho_d(\w, V) = \rho_d(\w)$ in Eq. (\ref{eq:ss-current})
neglecting its voltage dependency.  The latter becomes asymptotically
exact for $V\to 0$, defining the linear response regime. Deviations
from these curves are caused by the voltage dependency of the true
non-equilibrium spectral function.
The results are depicted in Fig.\ \ref{fig7:equi-nonequi} using the
SNRG data of Fig.\ \ref{fig5-SNRG-temperatures} for $T=\Gamma_0$.  The
SNRG curves (straight line) coincide with the equilibrium calculation
(dashed line) in the linear response regime, i.~e.\ $|eV|\ll
\Gamma_0$. However, the larger the electron-phonon coupling, the
smaller the validity range of the linear response regime.  Already at
very small finite voltages, we observe deviations from the I(V)
generated by the equilibrium $\rho_d(\w)$.  The excellent agreement
between the ISPI and the SNRG for $\lph/\Gamma_0=4$ results in the
small voltage regime -- see Fig.\ \ref{fig4-ISPI-SNRG} -- clearly
demonstrates that the SNRG correctly accounts for the bias dependence
of the spectral function.

\section{Conclusion}

We have applied the scattering states numerical renormalization group
approach to the charge-transport through a symmetric molecular
junction.  Since we have focused on the influence of a vibronic mode
on the transport, we have restricted ourselves to the investigation of
the spinless Anderson-Holstein Model.  We have started with a brief
review of the different regimes of the model and have connected them
to the polaronic energy shift $E_p$.

To set the stage for the non-equilibrium steady state currents we have
performed equilibrium calculations and have analysed the equilibrium
spectral functions in the different regimes.  We have demonstrated the
Franck-Condon blockade in the $I(V)$ curves found in the particle-hole
asymmetric case: the current is increasingly suppressed with
increasing electron-phonon coupling.

We have shown the temperature evolution of the $I(V)$ for two
different moderate temperatures to make contact to the ISPI approach
\cite{HutzenEgger2012}. While the ISPI is limited to large
temperatures due to the discretion of the memory kernel, the SNRG can
access arbitrarily low temperatures, the quantum coherence dominate
the transport properties at low temperatures.

At small voltages and strong electron-phonon coupling $\lph/\Gamma_0 >
2.5$ the shape change of the non-equilibrium spectral function leads
to a suppression of the current when the temperature is increased.
The temperature dependency of the current is governed by the
Fermi-functions of the lead for large voltages or small couplings
$\lph$.  We have shown that our non-equilibrium currents approach the
linear response regime for small voltages in which the voltage
dependency of the spectral function can be neglected.  With increasing
$\lph$, however, the validity radius of the linear response regime
becomes very small.

\section{Acknowledgments}
This paper is dedicated to the memory of Avi Schiller. He was not only
a dear friend but a collaborator for over 17 years and codeveloped
\cite{AndersSchiller2006,AndersSchiller2006} the non-equilibrium
extension of Wilson's numerical renormalization group which is the
foundation of the scattering states approach to steady state currents
applied in this paper.  Furthermore, we had many fruitful discussion
with Avi on the electron phonon coupling and profited a lot from his
exceptionally clear written paper \cite{EidelsteinSchiller2012}.
We acknowledges financial support by the German-Israel Foundation
through Grant No.\ 1035-36.14 and supercomputer support by the NIC,
Forschungszentrum J\"ulich under project no.\ HHB000.


%

\end{document}